\documentclass{osa-article}

\journal{osac}

\articletype{Research Article}

\begin{document}

\title{To What Extent Can Space Be Compressed? Bandwidth Limits of Spaceplates}

\author{Kunal Shastri\authormark{1}, Orad Reshef\authormark{2}, Robert W. Boyd\authormark{2,3,4}, Jeff S. Lundeen\authormark{2}, Francesco Monticone,\authormark{1,*}}

\address{\authormark{1}School of Electrical and Computer Engineering, Cornell University, Ithaca, NY 14853, USA, \authormark{2}Department of Physics, University of Ottawa, Ottawa, ON, Canada, \authormark{3}School of Electrical Engineering and Computer Science, University of Ottawa, Ottawa, ON, Canada, \authormark{4}Institute of Optics and Department of Physics and Astronomy, University of Rochester, Rochester, NY, USA}

\email{\authormark{*} francesco.monticone@cornell.edu}


\begin{abstract}
Spaceplates are novel flat-optic devices that implement the optical response of a free-space volume over a smaller length, effectively ``compressing space'' for light propagation. Together with flat lenses such as metalenses or diffractive lenses, spaceplates have the potential to enable a drastic miniaturization of any free-space optical system. While the fundamental and practical bounds on the performance metrics of flat lenses have been well studied in recent years, a similar understanding of the ultimate limits of spaceplates is lacking, especially regarding the issue of bandwidth, which remains as a crucial roadblock for the adoption of this platform. In this work, we derive fundamental bounds on the bandwidth of spaceplates as a function of their numerical aperture and compression ratio (ratio by which the free-space pathway is compressed). The general form of these bounds is universal and can be applied and specialized for different broad classes of space-compression devices, regardless of their particular implementation. Our findings also offer relevant insights into the physical mechanism at the origin of generic space-compression effects, and may guide the design of higher performance spaceplates, opening new opportunities for ultra-compact, monolithic, planar optical systems for a variety of applications.
\end{abstract}

\section{Introduction}
Optical systems, such as cameras, microscopes, telescopes, etc., typically have three major components: lenses, detectors, and free space. Metalenses \cite{ yu_flat_2014, khorasaninejad_polarization-insensitive_2016, liang_metalenses_2018, chen_broadband_2018, shrestha_broadband_2018, engelberg_advantages_2020} and diffractive lenses \cite{faklis_spectral_1995, sweeney_harmonic_1995, kim_increased_2013, meem_full-color_2018, banerji_imaging_2019} have the potential to replace conventional refractive lenses (within certain limits \cite{presutti_focusing_2020,engelberg_achromatic_2021, liang_high_2019}) and therefore, owing to their light weight and small form factor, contribute to the miniaturization of a multitude of optical systems. However, in this quest toward miniaturization, an often overlooked aspect of optical systems is the large free-space volumes between the detector and the lens, or between lenses, which is essential to allow light to acquire a distance-dependent and angle-dependent phase and achieve, for example, focusing at a certain distance. While free space obviously does not add to the overall weight, it can make dramatic contributions to the overall length of the system, thereby rendering the miniaturization of complex optical setups a challenge. 

An innovative solution to this problem was recently proposed by Reshef et al. \cite{reshef_optic_2021} by showing that a low-index isotropic or uniaxial slab, or a nonlocal, i.e., angle-dependent, (meta)material, can implement an optical transfer function approximating that of a volume of free space (or the selected background material) over a smaller distance. Such a device, named a ``spaceplate,'' can help shorten or even, in theory, fully replace the empty spaces in optical systems by effectively ``compressing'' space for light propagation. Several of us recently experimentally implemented and tested two broadband proof-of-concept spaceplates with a frequency window spanning the entire visible range, a first one, using a slab made of a uniaxial crystal, and a second one, using a glass cell containing air, both submerged in a higher index medium (oil). The compression ratio implemented by these homogeneous-material spaceplates, defined as the length of replaced free space (effective length of light propagation) over the actual spaceplate thickness, was however limited to less than $1.5$. In contrast, another pioneering work, by Guo et al. \cite{guo_squeeze_2020}, proposed a nonlocal spaceplate design leveraging the dispersion of a guided-mode resonance in a photonic crystal slab to implement the required transfer function, achieving a much larger compression ratio of $144$. The sharp line-width of the guided-mode resonance, however, limited the fractional spectral bandwidth of the device to $10^{-4}$ and the numerical aperture (NA) to $0.01$. In the same work, the authors proposed another photonic crystal-based spaceplate with a moderately larger NA of $0.11$; however, this was obtained at the cost of the compression ratio, which was reduced to $11.2$.

These early designs and demonstrations suggest that the operating bandwidth, compression ratio, and numerical aperture of spaceplates cannot be improved independently but are likely to be constrained by physical bounds and tradeoffs. This was further corroborated by the results of Ref. \cite{page_designing_2022}, where the three of the present authors observed a distinct tradeoff between compression ratio and NA in monochromatic multi-layered spaceplates designed using inverse design techniques, confirming earlier theoretical predictions by Chen and Monticone in Ref. \cite{chen_dielectric_2021}. 
Within this context, in this paper we first elucidate the basic physical mechanisms at the origin of generic space-compression effects, and we then derive physical bounds and tradeoffs on the performance of spaceplates, with a particular focus on the issue of bandwidth, which has not yet been addressed in the literature and remains as a crucial roadblock for the adoption of this platform. 
Specifically, starting from well-established limits on the delay-bandwidth product of linear time-invariant systems, we derive general fundamental bounds on the maximum achievable bandwidth of spaceplates, as a function of their compression ratio and numerical aperture. We then apply these theoretical results to answer important questions on the performance of ideal spaceplates, such as how optimal different spaceplate designs are, how to improve them, how the bandwidth limits of spaceplates and metalenses compare, and how high the maximum compression ratio and NA can be for an ideal spaceplate targeting the entire visible range.

\begin{figure}
	\centering\includegraphics[width=0.7\columnwidth]{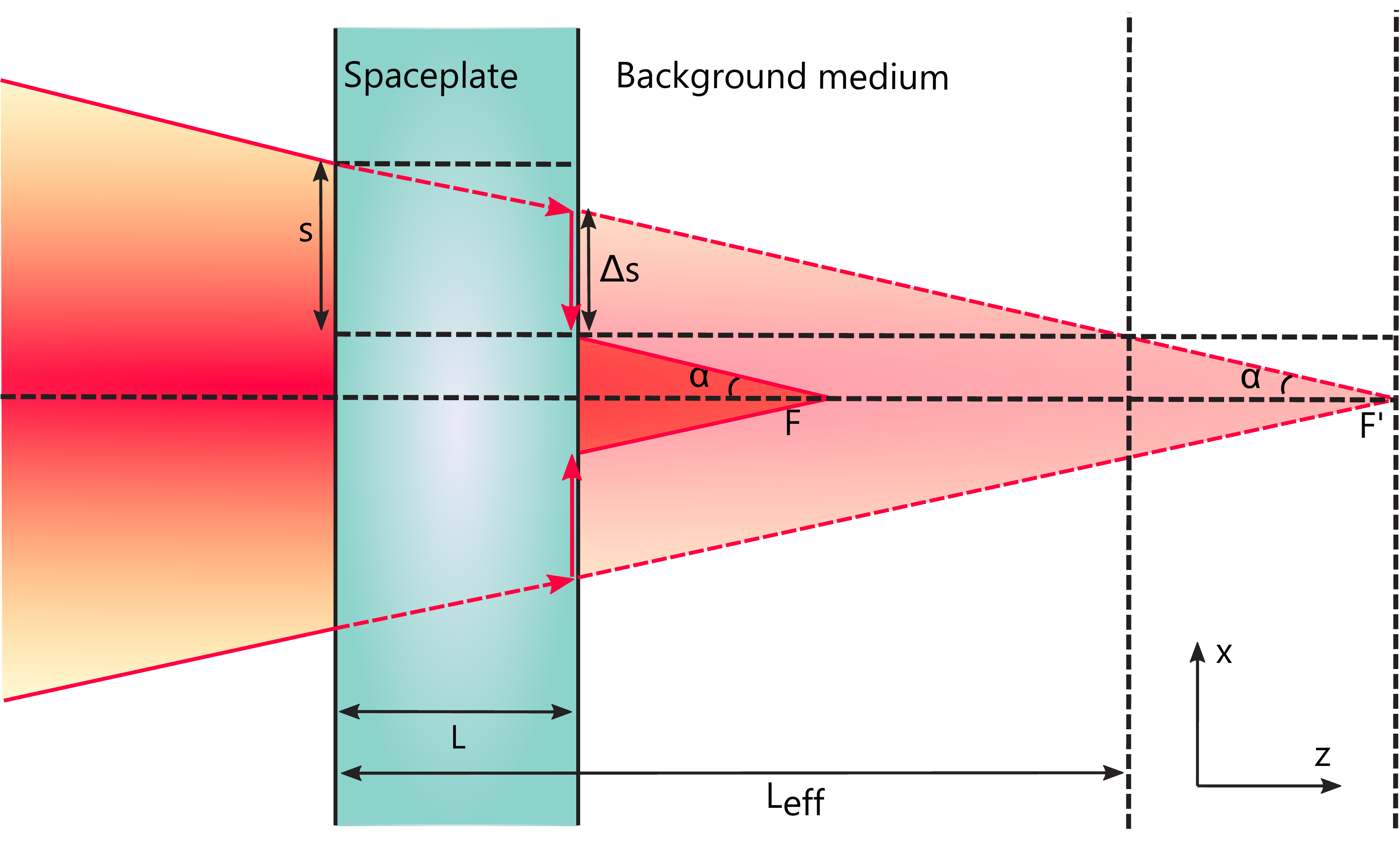}
	\caption{Comparison of a light beam converging at $F'$ in vacuum (or a background material) and at $F$ in the presence of an ideal spaceplate of length $L$. The spaceplate can effectively replace a region of space of length $L_{\textrm{eff}}$ by imparting an angle-dependent transverse shift, $\Delta s$, to propagating plane waves.} \label{fig:1}
\end{figure}

\section{Physical mechanism and fundamental limits}
Consider a light beam, perhaps shaped by a (meta)lens, that is converging/focusing at a distance $F'$ in vacuum (or a background material) and at $F$ in the presence of an ideal spaceplate, as illustrated in Fig. \ref{fig:1}. Intuitively, to mimic light traversing a distance $L_{\textrm{eff}}$ in a smaller space of length $L$ without distortions, obliquely incident plane waves should bend away from the surface normal inside the spaceplate, with a deflection angle dependent on the incident angle, and then re-emerge from the spaceplate propagating in the original direction. 
Crucially, to really mimic free space, the spaceplate in Fig. \ref{fig:1} should not reduce the focal distance through a ``lensing'' mechanism, i.e., deflecting the incident wave due to a position-dependent phase profile. Instead, the device should be \emph{transversely homogeneous}, not adding any focusing power, but simply implementing the transfer function of free space over a shorter distance. As mentioned in the Introduction, while this effect can be achieved with homogeneous media in a higher-index background, a promising way to realize space compression in free space is by using a suitable nonlocal structure, namely, a structure with an angle-dependent (i.e., transverse-momentum dependent) but position-independent response designed to match the angle-dependent phase response of free space. Structures with a nonlocal response, a property also known as \emph{spatial dispersion} -- the spatial analogue of the concept of frequency dispersion -- can be implemented, for instance, in the form of photonic-crystal slabs \cite{guo_squeeze_2020,long_polarization-independent_2021} or multilayer thin-film structures \cite{reshef_optic_2021,chen_dielectric_2021, gerken_multilayer_2003, silva_performing_2014}. 

Regardless of the implementation, it is clear that, as illustrated in Fig. \ref{fig:1}, a generic device for space compression needs to impart an angle-dependent transverse shift to the incident waves, \emph{relative to propagation in the background}, equal to $\Delta s=(L_{\textrm{eff}}-L)\tan(\alpha)$, where $\alpha$ is the incident angle with respect to the surface normal. (Interestingly, this is the dual process of a lens: whereas a lens is a position-dependent device that changes the angle of a light ray, a spaceplate is an angle-dependent device that changes the transverse spatial position of a light ray). 
In general, the transverse shift $s$ that a wave undergoes as it propagates through a material or structure is given by $s= v_{gx} \tau$ \cite{gerken_multilayer_2003}, where $v_{gx}$ is the transverse group velocity and $\tau$ is the total time delay (group delay) imparted by the material or structure, i.e., the difference, assumed here independent of the transverse position $x_0$, between the time the wave exits the structure at $(x,z)=(x_0+s,L)$ and the time it entered at $(x_0,0)$. (One could write $\tau=L/v_{gz}$, but it is important to note that for non-homogeneous or non-periodic structures, the group velocities $v_{gx}$ and $v_{gz}$ are effective quantities, representing the total effect of the structure, but are not constant within the structure itself \cite{gerken_limits_2005}). 
Then, with reference to Fig. \ref{fig:1}, the additional transverse shift with respect to the natural transverse displacement experienced by a wave propagating in the background medium can be written as $\Delta s= v_{gx} \tau - v_{gx}^0 \tau^0=v_{gx} \tau - c \sin(\alpha) \tau^0$, where $\tau^0=L / (c \cos(\alpha))$ is the time delay experienced by the wave in the background for the same distance $L$ along the $z$-axis, and $c=c_0/n_b$, is the wave velocity in the background medium, with refractive index $n_b$ assumed dispersionless. 
Thus, it is clear that in order to create a transverse displacement greater than the one in the background, and hence a space-compression effect, there are only two possible options: (i) By increasing the \emph{transverse group velocity} $v_{gx}$. This may be possible by simply re-directing waves away from the surface normal, for example through refraction if the refractive index of the spaceplate is less than in the surrounding medium, or through diffraction effects in transversely inhomogenous structures. The transverse group velocity $v_{gx}$ however can only be increased at most to approach $c_0$ and, in most cases of interest, such as dielectric stacks, the effective $v_{gx}$ is actually reduced, not increased, as further discussed below. (ii) By increasing the total \emph{time delay} with respect to propagation in the background, which implies that the spaceplate should act as an angle-dependent slow-light device. However, for a fixed length $L$ and other general properties of the structure, there are strict bounds on the maximum ``excess'' time delay, $\Delta T=\tau-\tau^0$ , that the structure can impart to a signal of certain bandwidth. Indeed, upper bounds on the product of delay and bandwidth will be crucial to determine the ultimate bandwidth limitations of a spaceplate.

Starting from the above equation for $\Delta s$, we can easily find a formula for the excess time delay that is required to implement a desired transverse displacement relative to propagation in the background,
\begin{equation} \label{eq:delay1}
	\Delta T = \frac{\Delta s + (\sin(\alpha)-v_{gx}/c)L \sec(\alpha)}{v_{gx}},
\end{equation} 
which means that a non-zero positive delay $\Delta T$ may be needed to realize a specified displacement $\Delta s$. The required $\Delta T$ may actually vanish for a sufficiently large $v_{gx}$, meaning that an increase in transverse velocity would be sufficient to achieve the desired displacement and, hence, space-compression effect. In many cases, however, especially for large $\alpha$ and small length $L$, it is clear that a positive $\Delta T$ would always be needed for a specified $\Delta s$ even for very large values of $v_{gx}$. Also note that if the desired $\Delta s$ were zero, and $v_{gx} > v_{gx}^0 = c \sin(\alpha)$, Eq. (\ref{eq:delay1}) would correctly predict that a negative $\Delta T$, i.e., a time advance, would be needed to achieve zero displacement; however, this is not of interest for our purposes and, therefore, for the spaceplate problem, we only consider positive or zero values of $\Delta T$, that is, the required time delay should be written as $\Delta T = \max[0,(\Delta s + (\sin(\alpha)-v_{gx}/c)L \sec(\alpha))/v_{gx}]$. 

If $\alpha_m$ is the maximum angular range of the spaceplate, an ideal spaceplate must be able to impart a time delay given by Eq. (\ref{eq:delay1}) for a lateral displacement $\Delta s_m=(L_{\textrm{eff}}-L)\tan(\alpha_m)$, which is the maximum displacement realized by the space-compression device. 
Using the definition of compression ratio, $R=L_{\textrm{eff}}/L$, and of numerical aperture, $\textrm{NA}=n_b\sin(\alpha_m)$, the required spaceplate group delay can be written as $\Delta T = L \max[(R\cdot\textrm{NA}/n_b - v_{gx}/c),0] / (v_{gx} \sqrt{1-(\textrm{NA}/n_b)^2})$. 
As mentioned above, whether this required $\Delta T$ can be implemented depends on the desired operating bandwidth $\Delta \omega$ of the spaceplate, namely, the maximum bandwidth of the fields the device is designed to interact with. In particular, for any linear time-invariant system, there are fundamental bounds on the delay-bandwidth product that can be achieved by the system \cite{miller_fundamental_2007-1}, which can generally be written in the form $\Delta T\Delta\omega\leq \kappa$, where $\kappa$ depends on some general properties of the structure, as further discussed below. Using the required $\Delta T$ derived above, a general expression for the bandwidth upper bound of any spaceplate can therefore be written as
\begin{equation} \label{eq:bound}
	\frac{\Delta\omega}{\omega_c}\leq\frac{1}{2\pi}\frac{\kappa}{L/\lambda_c}\frac{\sqrt{1-(\textrm{NA}/n_b)^2}~v_{gx}/c}{\max[(R\cdot\textrm{NA}/n_b - v_{gx}/c),0]},
\end{equation} 
where $\lambda_c$ is the center wavelength, in the background medium, and $\omega_c=2\pi f_c$ is the corresponding center angular frequency. Eq. (\ref{eq:bound}) is a key result of this paper, as it represents the general form of the bandwidth bound for any spaceplate, as a function of its numerical aperture and compression ratio, and it provides relevant quantitative insight into how easy or difficult it is to realize a space-compression effect. For example, the bound predicts the intuitive result that the fractional bandwidth reduces with increasing $R$ and goes to zero if $ R \rightarrow +\infty $. The maximum bandwidth also typically narrows with increasing NA, but the specific dependence on NA requires knowledge of $\kappa$, which may depend on the angle of incidence, as we will see in the following. 
Moreover, consistent with our discussion above, whenever the factor $(R\cdot\textrm{NA}/n_b - v_{gx}/c)$ vanishes or becomes negative for a certain combination of parameters, it is an indication that an increase in $v_{gx}$ with respect to the background-propagation case is sufficient to realize a transverse displacement and a space-compression effect without the need for the structure to impart an excess time delay and, therefore, without any bandwidth limitation due to delay-bandwidth constraints. This observation also offers important insight towards the design of optimal spaceplates, specifically about whether a structure acting as a slow-light device is useful or not. Indeed, it can easily be verified that if a spaceplate needs to be designed with a compression ratio and numerical aperture such that $R\cdot\textrm{NA} > n_b^2$, then a design based purely on increasing $v_{gx}$ (e.g., refraction in a low-index medium \cite{reshef_optic_2021}) is insufficient to achieve the desired space-compression effect, even in a scenario with $v_{gx} \rightarrow c_0$. Instead, a structure imparting an excess time delay is needed, which implies a finite bandwidth according to Eq. (\ref{eq:bound}). 

\begin{figure}
	\centering\includegraphics[width=0.85\columnwidth]{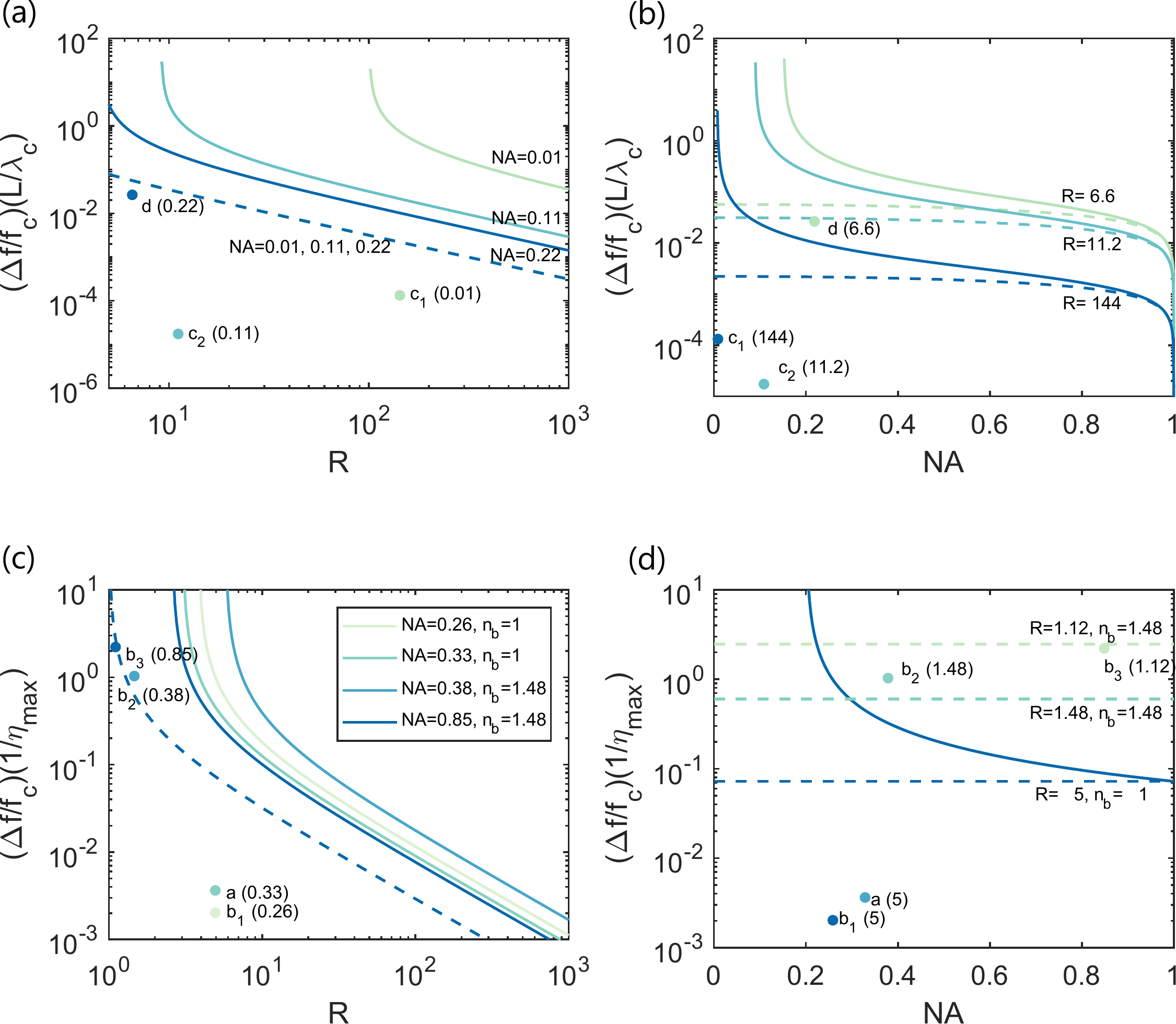}
	\caption{Bandwidth bounds compared with published spaceplate designs. The lines are the fractional bandwidth bounds derived in the main text (suitably normalized to allow for comparison between different designs) with solid lines representing the most general and loosest form of the bound with $v_{gx} \rightarrow c_0$, and the corresponding dashed lines are for a tighter version of the bound assuming $v_{gx}=c\sin(\alpha_m)=c\textrm{NA}/n_b$ (details in the main text). The dots in all plots represent the published designs given in Table 1, color-coded to match their respective bound curve. The bounds in (a) and (b) are calculated using the delay-bandwidth product for a single-mode resonator (applicable to spaceplates based on a single guided-mode resonance) and (c), (d) using the more general limit for the delay-bandwidth product of transversely invariant structures, corresponding to Eq. (\ref{eq:bound2}). (a),(c) Bandwidth bounds plotted as a function of the compression ratio $R$, for different values of the numerical aperture NA and background refractive index $n_b$ chosen on the basis of published designs. The number in parentheses next to the published design is the corresponding NA and $n_b$ reported. (b),(d) Bandwidth bounds plotted as a function of NA for different values of $R$ and $n_b$ chosen on the basis of published designs. The reported compression ratio for each published design is written in parentheses. If not specified, the background material is free space. Note that in panel (d) only one solid curve is plotted since, for the other two cases (corresponding to designs $b_2$ and $b_3$), the loosest bound is actually infinite as the reported parameters satisfy $R\cdot\textrm{NA}<n_b^2$ for any value of NA (hence, the space-compression effect can be achieved with no excess time delay; see discussion in the main text).} \label{fig:2}
\end{figure}

The case with the highest possible transverse velocity, $v_{gx} \rightarrow c_0$, is interesting as it provides the most general (and loosest) form of a bandwidth bound, valid for any structure regardless of their implementation. However, in realistic structures, $v_{gx}$ is usually significantly lower, angle-dependent, and it may be frequency dispersive. 
If better estimates of $v_{gx}$ for specific classes of structures are known, this may be plugged into Eq. (\ref{eq:bound}) to produce tighter limits on the bandwidth. For example, in dielectric multilayer thin-film structures, it was demonstrated in Ref. \cite{gerken_limits_2005} that $v_{gx}$ is approximately frequency-independent and can be estimated as $v_{gx}=c \sin(\alpha)/n_{\textrm{eff}}^2$, where $n_{\textrm{eff}}^2$ is a weighted average of the square of the refractive indices in the structure. In the following, we consider and plot two different versions of the bandwidth bound in Eq. (\ref{eq:bound}): the upper bound for the most optimistic, albeit unrealistic, scenario with $v_{gx}\rightarrow c_0$ (solid lines in Fig. \ref{fig:2}), and a tighter bound assuming that the transverse group velocity in the spaceplate does not exceed $v_{gx}$ in the background material for the considered maximum incidence angle, i.e., $v_{gx}=c \sin(\alpha_m)$ (dashed lines in Fig. \ref{fig:2}), which is relevant for spaceplates made of dielectric slabs/films.

\section{Specific cases}

The general bandwidth limit in Eq. (\ref{eq:bound}) can be specialized for any class of structures for which an upper bound $\kappa$ on the delay-bandwidth product is known. In the following, we consider two main types of spaceplates, based on different structures for which different expressions for $\kappa$ should be used, leading to different bandwidth limits. The first class of spaceplates are those based on structures supporting a guided-mode resonance, such as the photonic crystal-based spaceplate in Ref. \cite{guo_squeeze_2020,long_polarization-independent_2021} or the spaceplates based on a single Fabry-Perot cavity in Refs. \cite{chen_dielectric_2021} and \cite{mrnka_space-squeezing_2021}. In this case, assuming that only one resonance is used (as done in \cite{guo_squeeze_2020,mrnka_space-squeezing_2021}, and the first part of \cite{chen_dielectric_2021}), we can apply the delay-bandwidth product for a single-mode resonator, $\kappa = 2$ (valid for any individual single-mode resonator, reciprocal or nonreciprocal \cite{mann_nonreciprocal_2019}) to calculate the maximum achievable fractional bandwidth from Eq. (\ref{eq:bound}). The upper bound is plotted in Figs. \ref{fig:2}(a) and (b) (for both scenarios described above, $v_{gx} \rightarrow c_0$ and $v_{gx}=c \sin(\alpha_m)$) and compared with the spaceplate designs reported in the literature. As mentioned above, it is apparent from these plots that the bandwidth bound is lower for spaceplates with a higher NA or compression ratio. It can also be seen that the guided-mode resonance-based designs from \cite{guo_squeeze_2020}, marked as $c_1$ and $c_2$, obey the bound and, moreover, there is still significant room for improvement especially if a design could be found that not only imparts a delay, but also somewhat increases $v_{gx}$. The microwave design in \cite{mrnka_space-squeezing_2021}, marked as $d$, is the closest to the tighter bound (dashed line), albeit for a lower compression ratio, which is not surprising since this spaceplate structure is mostly empty, except for thin metallic metamaterial mirrors, hence the average $v_{gx}$ in the spaceplate is close to its free-space value. In addition, we note that, even at a single frequency, these resonator-based designs are constrained by other fundamental tradeoffs between the numerical aperture and the length of compressed space, as demonstrated in Ref. \cite{chen_dielectric_2021}.

The bandwidth bounds in Figs.~\ref{fig:2}(a) and (b) only apply to structures where the space-compression effect is based on a single resonance. Better performance can be achieved by including more resonators within the spaceplate thickness, increasing the number of available modes in the bandwidth of operation, as commonly done for slow-light devices. A more general limit to the delay-bandwidth product, valid for any linear time-invariant one-dimensional structure, was derived by Miller in \cite{miller_fundamental_2007, miller_fundamental_2007-1} (hereafter referred to as Miller's limit), which is fundamentally based on the idea that the product scales with the number of modes. In this case, the upper bound is given by $\kappa=\frac{\pi}{\sqrt{3}} \frac{L}{\lambda_c}\eta_{\max}$, where $\eta_{\max}=|(\epsilon_{\max}-\epsilon_{\min})/\epsilon_b|$ is the maximum permittivity contrast at any point within the structure and at any frequency within the considered bandwidth \cite{miller_fundamental_2007-1}. Miller's limit, however, was derived for one-dimensional structures, while spaceplates are clearly not one-dimensional devices. We circumvent this problem by noting that most spaceplate designs reported in the literature are transversely invariant, such as the homogeneous or multi-layered structures in \cite{chen_dielectric_2021, page_designing_2022, reshef_optic_2021}. In this case, the problem is formally equivalent to a one-dimensional one, as the transverse wavenumber $k_x$ is invariant in such a structure and, for transverse-electric (TE) or transverse-magnetic (TM) incident plane waves, a one-dimensional effective wave equation can be derived that does not depend on the transverse direction (see, e.g., \cite{s_a_tretyakov_analytical_2003, ishimaru_waves_2017}). This wave equation applies to the transverse component of the fields, $U_y(x,z)=u(z)e^{-j k_x x}$, and, in the non-magnetic case, can be written in the following general form: $k_{z,b}^{-2}~ \partial^2 u(z)/ \partial z^2 + u(z)=-\eta(z,\omega,\alpha) u(z)$, where $k_{z,b}$ is the longitudinal wavenumber in the background medium. Miller's limit can be derived for any system that can be described by a wave equation of this form \cite{miller_fundamental_2007}, but, compared to the original case, here the background wavenumber $k_b$ is replaced by the longitudinal wavenumber, $k_{z,b}=k_{b} \cos(\alpha)$ and the permittivity-contrast factor becomes angle-dependent $\eta(z,\omega,\alpha)=\eta(z,\omega)/\cos^2(\alpha)$. In addition, only for the TE case we have $\eta(z,\omega)=(\epsilon(z)-\epsilon_b)/\epsilon_b$ as in the original paper, whereas the TM case is significantly more complicated as this factor also depends on spatial derivatives of the permittivity distribution \cite{ishimaru_waves_2017}. While it can easily be verified that the TM case is approximately equal to the TE case if the permittivity does not change too rapidly in space, this dependence on the spatial derivatives of $\epsilon(z)$ suggests the intriguing possibility of potentially relaxed bounds for the TM case, which will be the subject of future studies. Here, instead, considering that ideal spaceplates should work for both polarizations, we consider the more stringent TE case, for which Miller's limit can be rederived with the only difference being that the wavelength $\lambda$ in the background medium is replaced by the longitudinal wavelength $\lambda_z=2\pi/k_{z,b}=\lambda/\cos(\alpha)$ and the maximum permittivity contrast becomes $\eta_{\max}(\alpha)=\eta_{\max}/\cos^2(\alpha)$. 
Finally, the modified upper bound can be written as, $\kappa=\frac{\pi}{\sqrt{3}} \frac{L}{\lambda_c}\eta_{\max}/\cos(\alpha)$, which converges to the one-dimensional case for normal incidence and increases with the angle, suggesting that larger time delays may be possible, for the same bandwidth, at oblique incidence. 

Substituting this modified $\kappa$ in Eq. (\ref{eq:bound}), we obtain the following limit on the fractional bandwidth of the space-compression effect,

\begin{equation} \label{eq:bound2}
	\frac{\Delta\omega}{\omega_c}\leq\frac{\eta_{\max}}{2\sqrt{3}}\frac{v_{gx}/c}{\max[(R \cdot\textrm{NA}/n_b - v_{gx}/c),0]},
\end{equation}
which is valid for any spaceplate based on a transversely invariant structure, such as homogeneous, multilayer, or graded-index slabs, nonlocal metamaterials composed of thin-film stacks, etc. The bound is plotted in Figs.~\ref{fig:2}(c) and (d) (again for both $v_{gx} \rightarrow c_0$ and $v_{gx}=c \sin(\alpha_m)$). As in the previous case, the maximum achievable fractional bandwidth decreases significantly for spaceplates with larger compression ratios, whereas the dependence on NA is significantly reduced (and completely removed in the case with $v_{gx}=c \sin(\alpha_m)$) because the requirement of a larger delay at oblique incidence to achieve a space-compression effect is compensated by the fact that the delay-bandwidth product limit is larger at oblique incidence for this type of structures. 

All spaceplate designs reported in the literature are below the appropriate upper bounds. Interestingly, the two implementations with the broadest bandwidth -- the isotropic and uniaxial slabs in a higher-index background \cite{reshef_optic_2021} (denoted as $b_2$ and $b_3$ in the figure) -- are permitted to operate over such a broad bandwidth (one of them exceeding the dashed-line bound for the case with no increase in $v_{gx}$) because their compression ratio and/or NA are sufficiently small such that the spaceplate effect can be achieved even \emph{without} an excess time delay, but purely through light refraction in the low-index material. Indeed, if $v_{gx}$ increases sufficiently within the spaceplate, and $R \cdot \textrm{NA}$ is small, then the bandwidth bound given by Eq. (\ref{eq:bound2}) diverges, meaning that the spaceplate is not limited by delay-bandwidth constraints. 
However, low-index refraction effects can be nearly frequency-independent only if the low-index material forming the spaceplate is air or an approximately dispersionless dielectric embedded in a higher-index dielectric background. Instead, if the background medium is free space, a spaceplate of this type necessitates a material with a refractive index lower than unity, which is only possible over a limited bandwidth and with non-zero frequency dispersion due to causality and passivity constraints on the properties of physical passive materials \cite{LANDAU_1984}. 
In such a case, the spaceplate bandwidth would be limited by the unavoidable dispersion of the low-index material, which translates into a dispersive $v_{gx}$ and, in turn, a frequency-dependent space-compression effect. Finally, as mentioned above, if the desired $R\cdot \textrm{NA} > n_b^2$, which is the case for all other designs reported in the literature (see also Table 1), then an excess time delay is always necessary, and Eq. (\ref{eq:bound2}) always predicts a finite bandwidth bound even if all materials are perfectly dispersionless.

\begin{figure}
	\centering\includegraphics[width=0.75\columnwidth]{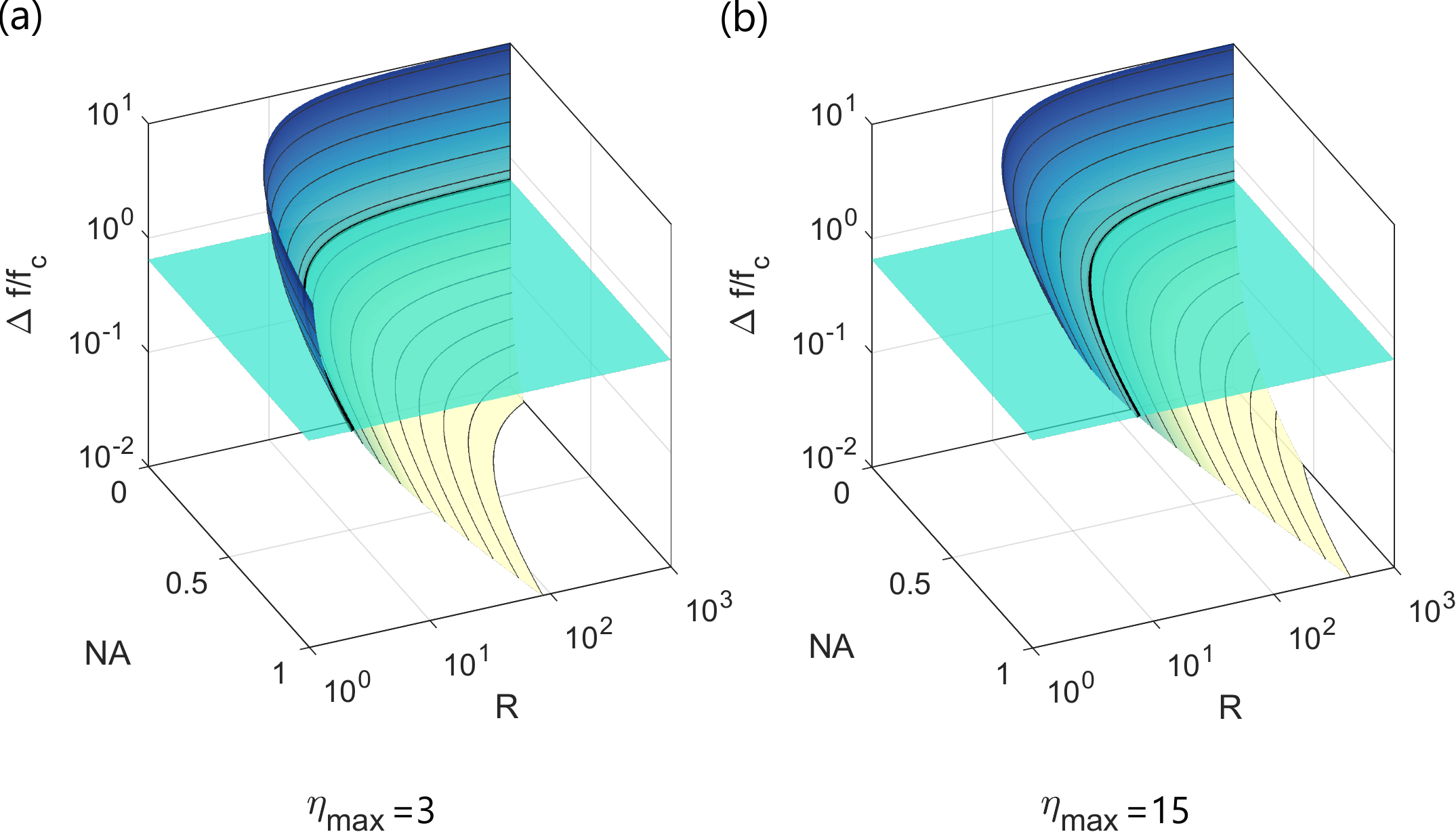}
	\caption{Illustration of the fractional bandwidth bound of a transversely homogeneous spaceplate as a function of its NA and compression ratio $R$, compared to the fractional bandwidth of the entire visible range (flat plane). The bound is calculated using Miller's delay-bandwidth product limit given by Eq. (\ref{eq:bound2}) (for the case $v_{gx} \rightarrow c_0$, leading to the loosest form of the bound) and with background refractive index set to unity. We considered two values of the maximum permittivity contrast: (a) $\eta_{\max}=3$ (low-contrast spaceplate) and (b) $\eta_{\max}=15$ (high-contrast spaceplate). The optimal trade-off between the NA and compression ratio for a fixed bandwidth is indicated by the solid lines (level curves), with the thicker line and flat plane corresponding to the fractional bandwidth of the visible range: $\Delta f/f_c$=0.65. } \label{fig:3}
\end{figure}

Since spaceplates based on multilayer structures have a large number of controllable variables such as layer thicknesses and refractive indices, they have the potential to access a wider range of spaceplate parameters. For example, it can be seen in Figs.~\ref{fig:2}(b),(d) that the multilayer structures denoted as $a$ \cite{chen_dielectric_2021} and $b_1$ \cite{reshef_optic_2021} achieve a good combination of relatively high NA, compression ratio, and bandwidth. At the same time, it is clear that there is still room for improvement, consistent with the fact that these early designs were not fully optimized. We anticipate that the inverse design techniques that have been recently developed to design high-performance multilayered spaceplates at a single frequency \cite{page_designing_2022} may be extended to also widen the operating bandwidth. Moreover, since the effective transverse group velocity in dielectric stacks was shown in Ref. \cite{gerken_limits_2005} to depend on a weighted average of the refractive indices in the structure, $v_{gx}=c \sin(\alpha)/n_{\textrm{eff}}^2$, we speculate that an ideal spaceplate design would utilize the highest possible refractive index to increase the delay-bandwidth product limit (which depends on the maximum permittivity contrast anywhere within the structure), while confining the high-index material in limited spatial regions (e.g., thin layers) such that the effective transverse group velocity would not decrease too much compared to free space. These qualitative criteria were satisfied for the spaceplates designed in Ref. \cite{chen_dielectric_2021}, which indeed show the highest fractional bandwidth of all the existing multilayer designs (see Table 1). Other more exotic options could perhaps involve combining both near-zero-index materials and dielectrics to increase the maximum time delay and $v_{gx}$ simultaneously.

\begin{table*}[]
	\resizebox{\columnwidth}{!}{\begin{tabular}{ l c c c c c c c c c}
		\hline

		Paper              &L/$\lambda_0$			&$\eta_{\max}$ 	&$NA$			&R		&R$_{max}$(Miller) 		&R$_{max}$($\kappa=2$) 	&$\frac{\Delta f}{f_c}$   	&$\frac{\Delta f}{f_c}\big\rvert_{max}$ (Miller) 		&$\frac{\Delta f}{f_c}\big\rvert_{max}$ ($\kappa=2$)\\
		
		\hline
		
		$a$\cite{chen_dielectric_2021} &9.00 &14.00 &0.33 &5.00 &81.83 &1.67 &0.05 &1.01 &8.35E-03 \\ 
		$b_{1}$\cite{reshef_optic_2021}(metamaterial) &6.56 &10.06 &0.26 &5.00 &146.20 &3.34 &0.02 &0.73 &0.01 \\ 
		$b_{2}$\cite{reshef_optic_2021}(low index material) &8.27E+03 &0.54 &0.38 &1.48 &1.29* &1.00* &0.55 &0.33* &7.75E-05* \\ 
		$b_{3}$\cite{reshef_optic_2021}(uniaxial crystal) &5.60E+04 &0.25 &0.85 &1.12 &1.13* &1.00* &0.55 &0.62* &3.98E-05* \\ 
		$c_1$\cite{guo_squeeze_2020}(small NA design) &1.22 &11.00 &0.01 &144.00 &3.02E+04 &2.49E+03 &1.05E-04 &0.02 &1.82E-03 \\ 
		$c_2$\cite{guo_squeeze_2020}(large NA design) &0.16 &11.00 &0.11 &11.20 &3.02E+04 &1.88E+04 &1.05E-04 &0.31 &0.19 \\ 
		$d$\cite{mrnka_space-squeezing_2021} &1.02 &--- &0.22 &6.60 &--- &13.13 &0.03 &--- &0.05 \\ 
		
		\hline
		
	\end{tabular}}
	\caption{List of various spaceplate designs reported in the literature and their thickness (L/$\lambda_c$), maximum permittivity contrast ($\eta_{\max}$), numerical aperture (NA), compression ratio ($R$), fractional bandwidth ($\Delta f/f_c$), and the corresponding maximum achievable compression ratio (for the reported NA) and maximum achievable bandwidth (for the reported $R$) calculated in this work. The limits are calculated using both the delay-bandwidth product for a single-mode resonator ($\kappa=2$) and the more general Miller's limit for transversely homogenous structures. For these calculations, we assumed $v_{gx}=c\sin(\alpha_m)=c\textrm{NA}/n_b$, namely, the transverse group velocity in the spaceplate is assumed not to exceed its value in the background material. This is true for all spaceplate designs reported in the literature, except the ones denoted with $b_2$ and $b_3$, which indeed approach or exceed this bound (numbers with an asterisk) since they are based on refraction in a low-index material, resulting in an increase in $v_{gx}$ (further details in the main text). Miller's bound is more relevant for $a$, $b_1$, $b_2$, $b_3$, whereas the single-resonance bound ($\kappa=2$) is more relevant for $c_1$, $c_2$, and $d$ ($\eta_{\max}$ and Miller's bound are not listed for $d$ as the corresponding spaceplate is made of very good conductors at microwave frequencies, hence $\eta_{\max}\rightarrow \infty$).} 
\end{table*}

Finally, for applications in optical imaging, it is interesting to assess the ultimate performance limits of spaceplates that are designed to function over the entire visible range. We do this by plotting in Fig. \ref{fig:3} the bandwidth bound for transversely homogeneous spaceplates, given by Eq. (\ref{eq:bound2}), as a function of NA and $R$, comparing it to the fractional bandwidth of the entire visible range. We consider two choices for the maximum permittivity contrast: $\eta_{\max}=3$ for a low-contrast spaceplate, and $\eta_{\max}=15$ for a high-contrast spaceplate (the latter is arguably the largest contrast that can be achieved at visible frequencies with natural materials \cite{shim_fundamental_2021}). 
As an example, considering a numerical aperture $\textrm{NA}=0.3$, an ideal spaceplate with $\eta_{\max}=3$ can achieve a compression ratio $R\approx 8$ over the entire visible spectrum, whereas a spaceplate with $\eta_{\max}=15$ can achieve a compression ratio of 25. These results demonstrate that, at least in theory, spaceplates with good performance in terms of numerical aperture and compression ratio can indeed cover the entire visible range.  

\begin{figure}
	\centering\includegraphics[width=0.8\columnwidth]{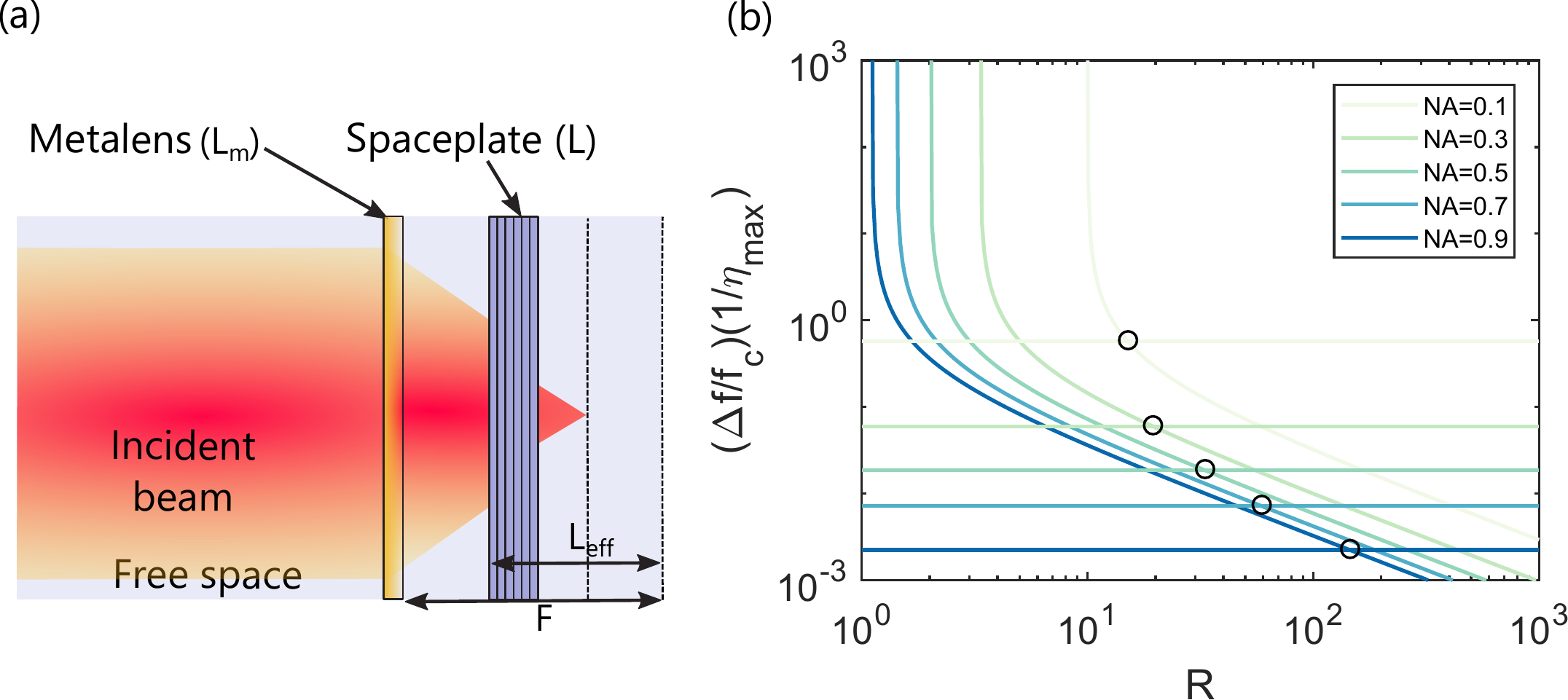}
	\caption{Comparison of bandwidth limits for metalenses and spaceplates. (a) Illustration of a flat optic focusing/imaging system consisting of a dispersion-engineered achromatic metalens followed by a spaceplate shifting the focal plane closer to the lens. (b) Normalized bandwidth bounds for a dispersion-engineered metalens with $F/L_m = 100$ (horizontal lines, Eq. (\ref{eq:metalens})) and a transversely invariant spaceplate with compression ratio $R$, for different values of NA. The spaceplate bound is calculated using Eq. (\ref{eq:bound2}) (for the case $v_{gx} \rightarrow c_0$, leading to the loosest form of the bound). The devices are assumed to be in free space ($n_b=1$) and to have the same maximum permittivity contrast $\eta_{\max}$. Crossing points between the bounds are denoted by black circles; for larger values of $R$ the spaceplate is the limiting factor to the bandwidth of the combined system, even in the most optimistic scenario.} \label{fig:4}
\end{figure}

\section{Cascaded metalens and spaceplate}

Cascading metalenses, spaceplates, and sensors can result in ultra-thin imaging platforms that are completely planar, ultra-compact, and potentially monolithic, which is one of the main motivations behind research on spaceplates and nonlocal flat optics \cite{reshef_optic_2021,guo_squeeze_2020,chen_dielectric_2021}. In this context, it is interesting and practically important to study whether the bandwidth would be limited by the spaceplate or by the flat lens itself. This is particularly relevant since one of the most popular classes of broadband flat lenses, namely, achromatic dispersion-engineered metalenses, are constrained by bandwidth limits analogous to the ones studied here as their operating mechanism is based on implementing, in addition to a standard phase profile, also a certain group delay profile to achieve dispersion compensation over a broad bandwidth \cite{chen_broadband_2018,shrestha_broadband_2018}. Bandwidth limits on achromatic metalenses, also originating from delay-bandwidth constraints, have been derived by Presutti and Monticone in Ref. \cite{presutti_focusing_2020}, where it was shown that the upper bandwidth bound is given by
\begin{equation} \label{eq:metalens}
	\frac{\Delta\omega}{\omega_c}\leq\frac{1}{2\sqrt{3}}\frac{L_m\eta_{\max}}{F}\frac{\sqrt{1-(NA/n_b)^2}}{1-\sqrt{1-(NA/n_b)^2}},
\end{equation} 
where $L_m$ and $F$ are the thickness and focal length of the metalens. We stress that, when a (meta)lens is followed by a spaceplate as illustrated in Fig. \ref{fig:4}(a), although focusing is achieved at a shorter distance due to the presence of a spaceplate, the focal length $F$ is unaltered since it is an intrinsic property of the lens itself (of its transverse profile) and it cannot be modified by the presence of a transversely homogeneous structure behind the lens. In other words, an ideal spaceplate does not change the imaging functionality or performance of the system, but it simply ``shifts'' the entire field distribution closer to the lens \cite{chen_dielectric_2021}. We also note that other classes of flat lenses, such as achromatic diffractive lenses \cite{banerji_imaging_2019,meem_broadband_2019} are not bound by the same limitations stemming from delay-bandwidth constraints since they are based on a different mechanism to achieve achromaticity, as nicely discussed in Ref. \cite{engelberg_achromatic_2021}; however, they suffer from other performance constraints, most notably the fact that they can provide near diffraction-limited performance only at very low Fresnel numbers \cite{engelberg_achromatic_2021}, making them less appealing for several applications. 

By comparing Eqs.~(\ref{eq:metalens}) and (\ref{eq:bound2}), we see that the bandwidth of a hybrid system composed of a metalens followed by a spaceplate, as in Fig. \ref{fig:4}(a), is limited by the space-compression effect if the compression ratio is sufficiently large. This is further confirmed in Fig. \ref{fig:4}(b), where we compare the two bounds, as functions of $R$ and for different values of NA, assuming $F/L_m = 100$ and the same maximum permittivity contrast for both devices. We see that, for this example, the crossing point is for compression ratios from ten to around one hundred, depending on the NA, above which, even in the most optimistic scenario, the spaceplate becomes the limiting factor with respect to bandwidth. These results also imply that even if flat lenses with broader bandwidths are employed, such as multilevel diffractive lenses \cite{banerji_extreme-depth--focus_2020,meem_broadband_2019}, the bandwidth of the combined system cannot improve if large compression ratios are desired.

\section{Conclusion}
 
In this paper, we have derived, for the first time, bandwidth limits on spaceplates, novel optical components that implement the optical response of free space over a shorter length. Using relevant upper bounds on the delay-bandwidth product of linear time-invariant structures, we have found that the space-compression effect is fundamentally limited in bandwidth if the necessary transverse spatial shift of a light beam propagating through the spaceplate requires the beam to be temporally delayed with respect to propagation in the background medium. In particular, we have shown that the achievable bandwidth necessarily decreases with an increase in the compression ratio. Importantly, however, we have found that the existing spaceplate designs are still relatively far from their performance limits, and that, at least in theory, compression ratios on the order of a few tens may be physically achievable over the entire visible range, using spaceplates made of realistic materials. While our results already apply to broad classes of spaceplates, we believe that future investigations should focus on identifying bandwidth bounds on more complex spaceplate devices, not necessarily transversely homogeneous or periodic, perhaps designed using inverse-design techniques applied to thick, volumetric, planar structures merging local metalenses and nonlocal spaceplate within the same volume \cite{lin_computational_2021}. Moreover, a better understanding of the fundamental limits to the transmission efficiency of spaceplates, and of the tradeoffs between efficiency, bandwidth and other metrics, would also be important to assess the potential of spaceplates for practical applications.

In summary, our work sheds new light on the physics, limitations, and potential of space-compression effects in optics and electromagnetics (some of these results and considerations may also be extended to other areas of wave physics). We believe these findings will help guide the design of new spaceplates with better performance, which may pave the way for the realization of ultra-compact, monolithic, planar optical systems for a variety of applications.

\begin{backmatter}

\bmsection{Funding}
F.M and K.S. acknowledge support from the National Science Foundation (NSF) with Grant No. 1741694, and the Air Force Office of Scientific Research with Grant No. FA9550-19-1-0043 through Dr. Arje Nachman. R.W.B. acknowledges support from the Natural Sciences and Engineering Research Council of Canada, the Canada Research Chairs program, and the Canada First Research Excellence Fund award on Transformative Quantum Technologies, US DARPA award W911NF-18-1-0369, US ARO award W911NF-18-1-0337, US Office of Naval Research MURI award N00014-20-1-2558, and DOE award GR530967. J.S.L acknowledges support from the Canada Research Chairs (CRC) program, the Natural Sciences and Engineering Research Council (NSERC), and the Canada First Research Excellence Fund award on Transformative Quantum Technologies.


\bmsection{Disclosures}
The authors declare no conflicts of interest.

\bmsection{Data availability} Data underlying the results presented in this paper are not publicly available at this time but may be obtained from the authors upon reasonable request.

\end{backmatter}


\bibliography{bibfile1}

\begin{thebibliography}{10}
\newcommand{\enquote}[1]{``#1''}

\bibitem{yu_flat_2014}
N.~Yu and F.~Capasso, \enquote{Flat optics with designer metasurfaces,}
  {\protect\JournalTitle{Nature Mater}} \textbf{13}, 139--150 (2014).

\bibitem{khorasaninejad_polarization-insensitive_2016}
M.~Khorasaninejad, A.~Y. Zhu, C.~Roques-Carmes, W.~T. Chen, J.~Oh, I.~Mishra,
  R.~C. Devlin, and F.~Capasso, \enquote{Polarization-{Insensitive}
  {Metalenses} at {Visible} {Wavelengths},} {\protect\JournalTitle{Nano Lett.}}
  \textbf{16}, 7229--7234 (2016).

\bibitem{liang_metalenses_2018}
Y.~Liang, Z.~Wei, J.~Guo, F.~Wang, H.~Meng, and H.~Liu, \enquote{Metalenses
  {Based} on {Symmetric} {Slab} {Waveguide} and c-{TiO2}: {Efficient}
  {Polarization}-{Insensitive} {Focusing} at {Visible} {Wavelengths},}
  {\protect\JournalTitle{Nanomaterials}} \textbf{8}, 699 (2018).

\bibitem{chen_broadband_2018}
W.~T. Chen, A.~Y. Zhu, V.~Sanjeev, M.~Khorasaninejad, Z.~Shi, E.~Lee, and
  F.~Capasso, \enquote{A broadband achromatic metalens for focusing and imaging
  in the visible,} {\protect\JournalTitle{Nature Nanotech}} \textbf{13},
  220--226 (2018).

\bibitem{shrestha_broadband_2018}
S.~Shrestha, A.~C. Overvig, M.~Lu, A.~Stein, and N.~Yu, \enquote{Broadband
  achromatic dielectric metalenses,} {\protect\JournalTitle{Light Sci Appl}}
  \textbf{7}, 85 (2018).

\bibitem{engelberg_advantages_2020}
J.~Engelberg and U.~Levy, \enquote{The advantages of metalenses over
  diffractive lenses,} {\protect\JournalTitle{Nat Commun}} \textbf{11}, 1991
  (2020).

\bibitem{faklis_spectral_1995}
D.~Faklis and G.~M. Morris, \enquote{Spectral properties of multiorder
  diffractive lenses,} {\protect\JournalTitle{Appl. Opt.}} \textbf{34},
  2462--2468 (1995).

\bibitem{sweeney_harmonic_1995}
D.~W. Sweeney and G.~E. Sommargren, \enquote{Harmonic diffractive lenses,}
  {\protect\JournalTitle{Appl. Opt.}} \textbf{34}, 2469--2475 (1995).

\bibitem{kim_increased_2013}
G.~Kim, J.~A. Dominguez-Caballero, H.~Lee, D.~J. Friedman, and R.~Menon,
  \enquote{Increased {Photovoltaic} {Power} {Output} via {Diffractive}
  {Spectrum} {Separation},} {\protect\JournalTitle{Phys. Rev. Lett.}}
  \textbf{110}, 123901 (2013).

\bibitem{meem_full-color_2018}
M.~Meem, A.~Majumder, and R.~Menon, \enquote{Full-color video and still imaging
  using two flat lenses,} {\protect\JournalTitle{Opt. Express}} \textbf{26},
  26866--26871 (2018).

\bibitem{banerji_imaging_2019}
S.~Banerji, M.~Meem, A.~Majumder, F.~G. Vasquez, B.~Sensale-Rodriguez, and
  R.~Menon, \enquote{Imaging with flat optics: metalenses or diffractive
  lenses?} {\protect\JournalTitle{Optica}} \textbf{6}, 805--810 (2019).

\bibitem{presutti_focusing_2020}
F.~Presutti and F.~Monticone, \enquote{Focusing on bandwidth: achromatic
  metalens limits,} {\protect\JournalTitle{Optica}} \textbf{7}, 624 (2020).

\bibitem{engelberg_achromatic_2021}
J.~Engelberg and U.~Levy, \enquote{Achromatic flat lens performance limits,}
  {\protect\JournalTitle{Optica}} \textbf{8}, 834--845 (2021).

\bibitem{liang_high_2019}
H.~Liang, A.~Martins, B.-H.~V. Borges, J.~Zhou, E.~R. Martins, J.~Li, and T.~F.
  Krauss, \enquote{High performance metalenses: numerical aperture,
  aberrations, chromaticity, and trade-offs,} {\protect\JournalTitle{Optica}}
  \textbf{6}, 1461--1470 (2019).

\bibitem{reshef_optic_2021}
O.~Reshef, M.~P. DelMastro, K.~K.~M. Bearne, A.~H. Alhulaymi, L.~Giner, R.~W.
  Boyd, and J.~S. Lundeen, \enquote{An optic to replace space and its
  application towards ultra-thin imaging systems,} {\protect\JournalTitle{Nat
  Commun}} \textbf{12}, 3512 (2021).

\bibitem{guo_squeeze_2020}
C.~Guo, H.~Wang, and S.~Fan, \enquote{Squeeze free space with nonlocal flat
  optics,} {\protect\JournalTitle{Optica}} \textbf{7}, 1133 (2020).

\bibitem{page_designing_2022}
J.~T.~R. Pagé, O.~Reshef, R.~W. Boyd, and J.~S. Lundeen, \enquote{Designing
  high-performance propagation-compressing spaceplates using thin-film
  multilayer stacks,} {\protect\JournalTitle{Opt. Express}} \textbf{30},
  2197--2205 (2022).

\bibitem{chen_dielectric_2021}
A.~Chen and F.~Monticone, \enquote{Dielectric {Nonlocal} {Metasurfaces} for
  {Fully} {Solid}-{State} {Ultrathin} {Optical} {Systems},}
  {\protect\JournalTitle{ACS Photonics}} \textbf{8}, 1439--1447 (2021).

\bibitem{long_polarization-independent_2021}
O.~Y. Long, C.~Guo, W.~Jin, and S.~Fan, \enquote{Polarization-independent
  isotropic nonlocal metasurfaces with wavelength-controlled functionality,}
  {\protect\JournalTitle{arXiv:2112.06731 [physics]}}  (2021).

\bibitem{gerken_multilayer_2003}
M.~Gerken and D.~A.~B. Miller, \enquote{Multilayer thin-film structures with
  high spatial dispersion,} {\protect\JournalTitle{Appl. Opt.}} \textbf{42},
  1330 (2003).

\bibitem{silva_performing_2014}
A.~Silva, F.~Monticone, G.~Castaldi, V.~Galdi, A.~Alù, and N.~Engheta,
  \enquote{Performing {Mathematical} {Operations} with {Metamaterials},}
  {\protect\JournalTitle{Science}} \textbf{343}, 160--163 (2014).

\bibitem{gerken_limits_2005}
M.~Gerken and D.~A.~B. Miller, \enquote{Limits on the performance of dispersive
  thin-film stacks,} {\protect\JournalTitle{Appl. Opt.}} \textbf{44},
  3349--3357 (2005).

\bibitem{miller_fundamental_2007-1}
D.~A.~B. Miller, \enquote{Fundamental {Limit} to {Linear} {One}-{Dimensional}
  {Slow} {Light} {Structures},} {\protect\JournalTitle{Phys. Rev. Lett.}}
  \textbf{99}, 203903 (2007).

\bibitem{mrnka_space-squeezing_2021}
M.~Mrnka, E.~Hendry, J.~Láčík, L.~E. Barr, and D.~B. Phillips,
  \enquote{Space-squeezing optics in the microwave spectral region,}
  {\protect\JournalTitle{arXiv:2110.15022 [physics]}}  (2021).

\bibitem{mann_nonreciprocal_2019}
S.~A. Mann, D.~L. Sounas, and A.~Alù, \enquote{Nonreciprocal cavities and the
  time–bandwidth limit,} {\protect\JournalTitle{Optica}} \textbf{6}, 104--110
  (2019).

\bibitem{miller_fundamental_2007}
D.~A.~B. Miller, \enquote{Fundamental limit for optical components,}
  {\protect\JournalTitle{J. Opt. Soc. Am. B}} \textbf{24}, A1 (2007).

\bibitem{s_a_tretyakov_analytical_2003}
S.~A. Tretyakov, \emph{Analytical {Modeling} in {Applied} {Electromagnetics}}
  (2003).

\bibitem{ishimaru_waves_2017}
A.~Ishimaru, \enquote{Waves in {Inhomogeneous} and {Layered} {Media},} in
  \emph{Electromagnetic {Wave} {Propagation}, {Radiation}, and {Scattering},}
  (John Wiley \& Sons, Ltd, 2017), pp. 35--84.

\bibitem{LANDAU_1984}
L.~Landau and E.~Lifshitz, vol.~8 of \emph{Course of Theoretical Physics}
  (Pergamon, Amsterdam, 1984), second edition ed.

\bibitem{shim_fundamental_2021}
H.~Shim, F.~Monticone, and O.~D. Miller, \enquote{Fundamental {Limits} to the
  {Refractive} {Index} of {Transparent} {Optical} {Materials},}
  {\protect\JournalTitle{Advanced Materials}} \textbf{33}, 2103946 (2021).

\bibitem{meem_broadband_2019}
M.~Meem, S.~Banerji, A.~Majumder, F.~G. Vasquez, B.~Sensale-Rodriguez, and
  R.~Menon, \enquote{Broadband lightweight flat lenses for long-wave infrared
  imaging,} {\protect\JournalTitle{PNAS}} \textbf{116}, 21375--21378 (2019).

\bibitem{banerji_extreme-depth--focus_2020}
S.~Banerji, M.~Meem, A.~Majumder, B.~Sensale-Rodriguez, and R.~Menon,
  \enquote{Extreme-depth-of-focus imaging with a flat lens,}
  {\protect\JournalTitle{Optica}} \textbf{7}, 214 (2020).

\bibitem{lin_computational_2021}
Z.~Lin, C.~Roques-Carmes, R.~E. Christiansen, M.~Soljačić, and S.~G. Johnson,
  \enquote{Computational inverse design for ultra-compact single-piece
  metalenses free of chromatic and angular aberration,}
  {\protect\JournalTitle{Appl. Phys. Lett.}} \textbf{118}, 041104 (2021).

\end{thebibliography}

\end{document}